\documentclass[unnumsec,webpdf,contemporary,large,numbered]{oup-authoring-template}
\usepackage{bm}
\newcommand{\societylogo}{}
\makeatletter
\let\ps@opening\ps@plain
\@numbibtrue
\usepackage{etoolbox}
\patchcmd{\@maketitle}{\selectfont\centering\@address}{\selectfont\raggedright\@address}{}{\PackageError{manuscript}{Affiliation alignment patch failed}{}}
\patchcmd{\@maketitle}{\selectfont\centering\@address}{\selectfont\raggedright\@address}{}{\PackageError{manuscript}{Affiliation alignment patch failed}{}}
\makeatother
\setcitestyle{numbers,square}
\AtBeginDocument{\global\advance\stockwidth by 15mm
  \global\advance\pdfpagewidth by 15mm
  \global\advance\hoffset by 15mm}
\begin{document}
\journaltitle{}
\DOI{}
\copyrightyear{}
\pubyear{}
\vol{}
\issue{}
\access{}
\appnotes{Research Article}
\firstpage{1}
\title[Fast Functional Clustering]{Fast Functional Clustering via Closed-Form Covariance Updates}
\author[1,2]{Yu Wang}
\author[3]{Jiaze Guo}
\author[1,2]{Yao Chen}
\author[4]{Tianxiang Hao}
\author[2,5,$\ast$]{Rongling Wu}
\address[1]{Institute of Statistics and Big Data, Renmin University of China, Beijing 100872, China}
\address[2]{Beijing Key Laboratory of Topological Statistics and Applications for Complex Systems, Beijing Institute of Mathematical Sciences and Applications, Beijing 101408, China}
\address[3]{School of Grassland Science, Beijing Forestry University, Beijing 100083, China}
\address[4]{The Tsinghua Shenzhen International Graduate School, Tsinghua University, Shenzhen, China}
\address[5]{Yau Mathematical Sciences Center, Tsinghua University, Beijing 100084, China}
\corresp[$\ast$]{Corresponding author: Rongling Wu.}
\abstract{Functional clustering identifies coordinated molecular
profiles, but repeated numerical covariance optimization can make
mixture-model fitting computationally demanding. We develop a fast
functional clustering procedure for joint analysis across conditions,
combining linear mean profiles on a standardized log niche coordinate
with first-order structured antedependence (SAD(1)) covariance.
Weighted residual statistics yield closed-form conditional covariance
updates under four parameter-sharing specifications, while an
innovation representation simplifies likelihood evaluation. In
same-R comparisons with numerical optimization of identical conditional
objectives, the closed-form solver was 91.5--201.4 times as fast,
with objective agreement in all 60 targets. Under matched complete
mixture fits, it achieved a 1.87--2.37-fold speedup with identical
cluster assignments. Recovery simulations showed higher module
classification accuracy and lower mean-curve error than $K$-means
and a Gaussian mixture with shared full covariance when the mean
structure was correctly specified. SAD(1) provided an additional
recovery gain under strong residual correlation, whereas curved-mean
controls identified the importance of an appropriate mean family.
Applications to single-condition gut microbiota, paired leaf--root
microbiota and three-visit human gut multi-omics illustrate the
framework across different numbers of conditions. Closed-form
covariance estimation makes joint functional clustering substantially
faster while retaining the fitted model's likelihood and assignments.}

\maketitle

\section{Introduction}
\label{sec:introduction}

Functional clustering groups molecular features by the shapes
of their expression profiles, providing a way to describe
coordinated variation across samples and experimental conditions.
In a model-based formulation, each cluster is represented by a
probability distribution, with a mean function describing its
characteristic profile and a covariance structure describing
residual dependence \citep{FraleyRaftery2002}. This combination
connects module membership to interpretable profile parameters.
It has been used to identify periodic gene-expression patterns
and to characterize expression responses across environments
\citep{KimEtAl2008,WangEtAl2012}.

The coordinate underlying a functional profile can also be
constructed from variation among samples. Niche-based approaches
order samples by an aggregate index of the system and relate
individual feature values to that index
\citep{ChenEtAl2019,DongEtAl2023}. Allometric mean functions
then describe how a feature varies with the overall state of
the system, extending functional clustering to profiles indexed
by sample-level variation \citep{PanEtAl2026}. For data collected
under several conditions, a joint analysis can group features
according to their combined profiles while retaining a distinct
mean function for each condition. This formulation provides a
common set of modules for comparing condition-specific patterns.

Efficient estimation of residual dependence is central to this
analysis. An unrestricted covariance matrix requires a number
of parameters that grows quadratically with profile length.
Antedependence models offer a parsimonious alternative by
expressing dependence through conditional regressions on earlier
ordered measurements \citep{Gabriel1962}. Structured
antedependence models have been used to represent covariance
in longitudinal and functional data
\citep{NunezAntonZimmerman2000,ZhaoEtAl2005}.
The first-order structured antedependence model, SAD(1), is
particularly attractive because a small number of parameters
describes both residual dependence and the resulting marginal
variance structure.

Parameter estimation, however, can require substantial
computation even with a compact covariance model.
Existing FunClu formulations have combined the
expectation--maximization (EM) algorithm with numerical
optimization of the mean and covariance parameters.
The framework of \citet{WangEtAl2012} embedded simplex
optimization within EM, and the allometric formulation of
\citet{PanEtAl2026} used gradient-based optimization for these
parameters. Such updates repeatedly evaluate the covariance
objective within each EM iteration. Multiple initializations
and comparisons of candidate cluster numbers multiply this
work. The conditional regression structure of SAD(1) offers
an opportunity to replace these repeated numerical searches
with direct parameter updates and to evaluate the likelihood
through residual innovations.

Here, we develop a functional clustering procedure with
closed-form SAD(1) conditional updates for joint profiles
across conditions. We model the mean as a linear function of
the standardized log niche coordinate and consider four
covariance specifications that share the antedependence
coefficient, the innovation variance, both or neither across
conditions. Within the expectation/conditional maximization
framework \citep{MengRubin1993}, estimation alternates
between weighted least-squares mean updates and direct
conditional covariance updates.
An innovation representation and cached quadratic statistics
reduce repeated likelihood calculations. We quantify the
computational gain through matched-model fitting, conditional
covariance solving and comparisons with the original FunClu
implementation. Closed-form covariance solving achieves
approximately two orders of magnitude of acceleration in the
same-R comparisons. We further demonstrate the analysis
using the published gut and paired leaf--root microbiota
examples distributed with idopNetwork \citep{DongEtAl2023},
followed by gut multi-omics profiles from 149 participants
measured at three visits \citep{ZhangEtAl2026}.

\section{Method}

\subsection{Mixture model}

We cluster features by their joint expression profiles across
$L$ conditions. Consider $p$ features measured in $n_\ell$
samples under condition $\ell$. Within each condition, samples
are ordered by their subtotal niche values, with corresponding
coordinates $\tau_1^{(\ell)}\leq\cdots\leq
\tau_{n_\ell}^{(\ell)}$. Let
$y_j^{(\ell)}(\tau_i^{(\ell)})$ denote the expression value
of feature $j$ in the $i$-th ordered sample. The observed
profile of feature $j$ under condition $\ell$ is
\[
\bm y_j^{(\ell)}
=
\bigl(
y_j^{(\ell)}(\tau_1^{(\ell)}),\ldots,
y_j^{(\ell)}(\tau_{n_\ell}^{(\ell)})
\bigr)^\top
\in\mathbb R^{n_\ell}.
\]
The niche coordinates may be unevenly spaced, and sample
sizes may differ across conditions. We concatenate the
condition-specific profiles into
$\bm y_j=\operatorname{col}\bigl(
\bm y_j^{(1)},\ldots,\bm y_j^{(L)}
\bigr)\in\mathbb R^N$,
where $\operatorname{col}(\cdot)$ stacks column vectors and
$N=\sum_{\ell=1}^{L}n_\ell$.

Following the functional clustering framework
\citep{WangEtAl2012,PanEtAl2026}, we represent distinct
expression patterns by $K$ Gaussian mixture components.
Let $C_j\in\{1,\ldots,K\}$ denote the latent cluster
membership of feature $j$, with $\Pr(C_j=k)=\pi_k$.
Each feature has a single cluster membership across all
conditions. Conditional on $C_j=k$, its joint profile follows
$\bm y_j\mid C_j=k\sim N_N(\bm\mu_k,\bm\Sigma)$.
The marginal density is therefore
\begin{equation}
    f(\bm y_j;\bm\Theta)
    =\sum_{k=1}^{K}\pi_k\,
    f_k(\bm y_j;\bm\mu_k,\bm\Sigma),
\end{equation}
where $\pi_k\geq0$, $\sum_{k=1}^{K}\pi_k=1$, and
$f_k$ is the $N$-variate normal density with mean
\begin{equation}
    \bm\mu_k
    =\operatorname{col}\bigl(
    \bm\mu_k^{(1)},\ldots,\bm\mu_k^{(L)}
    \bigr).
\end{equation}
Here,
$\bm\mu_k^{(\ell)}=
\bigl(\mu_k^{(\ell)}(\tau_1^{(\ell)}),\ldots,
\mu_k^{(\ell)}(\tau_{n_\ell}^{(\ell)})\bigr)^\top$
describes the mean profile of cluster $k$ under condition
$\ell$. Thus, a cluster is characterized by a set of
condition-specific mean profiles.

The covariance matrix $\bm\Sigma\in\mathbb R^{N\times N}$
is positive definite and shared by all clusters. We assume
that profiles from different conditions are independent
conditional on cluster membership, giving
\begin{equation}
    \bm\Sigma
    =\operatorname{blockdiag}\bigl(
    \bm\Sigma^{(1)},\ldots,\bm\Sigma^{(L)}
    \bigr),
\end{equation}
where $\bm\Sigma^{(\ell)}\in\mathbb R^{n_\ell\times n_\ell}$
captures residual dependence among the ordered samples
within condition $\ell$.

Treating the feature profiles as independent observations,
the observed-data likelihood is
\begin{equation}
    L(\bm\Theta\mid\bm Y)
    =\prod_{j=1}^{p}
    \sum_{k=1}^{K}\pi_k\,
    f_k(\bm y_j;\bm\mu_k,\bm\Sigma),
\end{equation}
where $\bm Y=(\bm y_1,\ldots,\bm y_p)$ and
$\bm\Theta=(\bm\pi,\bm\Theta_\mu,\bm\Theta_\Sigma)$
collects the unknown parameters. Here,
$\bm\pi=(\pi_1,\ldots,\pi_K)$ contains the mixture
proportions,
$\bm\Theta_\mu=
\{\bm\Theta_{\mu_k}^{(\ell)}:
k=1,\ldots,K;\ \ell=1,\ldots,L\}$
specifies the cluster- and condition-specific mean
functions, and $\bm\Theta_\Sigma$ specifies the common
covariance structure. The mean vectors and covariance
blocks are parameterized by the functions and dependence
models defined in the following subsections.

\subsection{Structural modeling of mean vectors}

Functional clustering characterizes each cluster through a
mean function whose parameters describe its expression
pattern \citep{WangEtAl2012}. For profiles indexed by the
subtotal niche, allometric scaling provides a functional
form for relating individual expression to the aggregate
expression of the system \citep{PanEtAl2026}. We use its
log-linear parameterization to model the mean profiles.
For condition $\ell$, define
\begin{equation}
    z_i^{(\ell)}
    =\frac{\ln\tau_i^{(\ell)}-m_\ell}{d_\ell},
    \qquad i=1,\ldots,n_\ell,
\end{equation}
where $m_\ell$ and $d_\ell>0$ are the mean and standard
deviation of the log niche coordinates, with
$d_\ell^2=n_\ell^{-1}\sum_{i=1}^{n_\ell}
(\ln\tau_i^{(\ell)}-m_\ell)^2$.
The mean profile of cluster $k$ under condition $\ell$ is
\begin{equation}
    \mu_k^{(\ell)}(\tau_i^{(\ell)})
    =\alpha_k^{(\ell)}
    +\beta_k^{(\ell)}z_i^{(\ell)}.
    \label{eq:mean-profile}
\end{equation}
Here, $\alpha_k^{(\ell)}$ is the mean level at the geometric
mean niche coordinate $\exp(m_\ell)$, and
$\beta_k^{(\ell)}$ describes the change in mean expression
per standard deviation of the log niche coordinate.
Positive and negative slopes describe increasing and
decreasing profiles, respectively, whereas
$\beta_k^{(\ell)}=0$ gives a constant profile.

The connection to allometric scaling is expressed by
\begin{equation}
\begin{gathered}
    \exp\{\mu_k^{(\ell)}(\tau)\}=a_k^{(\ell)}\tau^{b_k^{(\ell)}},\\
    b_k^{(\ell)}=\frac{\beta_k^{(\ell)}}{d_\ell},\qquad
    a_k^{(\ell)}=\exp\{\alpha_k^{(\ell)}-b_k^{(\ell)}m_\ell\}.
\end{gathered}
\end{equation}
Equation~\eqref{eq:mean-profile} specifies the Gaussian
mean on the response scale used in the mixture model.
The coefficients are allowed to differ among clusters
and conditions, so each cluster is described by $L$
mean profiles. Collecting their parameters gives
\begin{equation}
    \bm\Theta_\mu
    =\left\{
    \bm\Theta_{\mu_k}^{(\ell)}
    =(\alpha_k^{(\ell)},\beta_k^{(\ell)}):
    k=1,\ldots,K;\ \ell=1,\ldots,L
    \right\}.
\end{equation}
This specification uses $2KL$ mean parameters.
Initial coefficients are obtained by ordinary least
squares within provisional clusters
(Appendix~\ref{app:ols}); they are subsequently updated
under the structured covariance model within the EM
algorithm.

\subsection{Structural modeling of covariance structure}

The covariance structure describes the dependence among
expression measurements along the ordered niche sequence.
For feature $j$ belonging to cluster $k$, write
$\bm e_{jk}^{(\ell)}
=\bm y_j^{(\ell)}-\bm\mu_k^{(\ell)}$ for its residual
profile under condition $\ell$.
We model this dependence using a first-order structured
antedependence model, SAD(1)
\citep{Gabriel1962,NunezAntonZimmerman2000,ZhaoEtAl2005}:
\begin{equation}
\begin{aligned}
    e_{jk}^{(\ell)}(1)
    &=\varepsilon_{jk}^{(\ell)}(1),\\
    e_{jk}^{(\ell)}(i)
    &=\phi_\ell e_{jk}^{(\ell)}(i-1)
    +\varepsilon_{jk}^{(\ell)}(i),
    \qquad i=2,\ldots,n_\ell,
\end{aligned}
\label{eq:sad-recursion}
\end{equation}
where the innovations are independent with
$\varepsilon_{jk}^{(\ell)}(i)\sim N(0,\gamma_\ell^2)$,
$\gamma_\ell>0$ and $|\phi_\ell|<1$.
The antedependence coefficient $\phi_\ell$ describes the
residual dependence between successive ordered positions,
and $\gamma_\ell^2$ is the innovation variance.
Both parameters are constant within a condition and
shared by all clusters. Dependence is indexed by position
in the niche ordering, so successive positions need not
represent equal differences in niche values.

The recursion in equation~\eqref{eq:sad-recursion} gives
\begin{equation}
\begin{aligned}
    \Sigma_{ih}^{(\ell)}
    &=\operatorname{Cov}\bigl(e_{jk}^{(\ell)}(i),e_{jk}^{(\ell)}(h)\bigr)\\
    &=\gamma_\ell^2\phi_\ell^{|i-h|}
    \frac{1-\phi_\ell^{2\min(i,h)}}{1-\phi_\ell^2},
    \quad i,h=1,\ldots,n_\ell.
\end{aligned}
    \label{eq:sad-covariance}
\end{equation}
In particular, the marginal variance at position $i$ is
$\gamma_\ell^2(1-\phi_\ell^{2i})/(1-\phi_\ell^2)$.
Thus, the model accommodates position-dependent variances
and correlations using two parameters per condition.
When $\phi_\ell=0$, the covariance reduces to
$\gamma_\ell^2\bm I_{n_\ell}$.

For likelihood evaluation, let $\bm B_\ell$ be the
$n_\ell\times n_\ell$ lower bidiagonal matrix with diagonal
entries equal to one and subdiagonal entries equal to
$-\phi_\ell$. Then
$\bm B_\ell\bm e_{jk}^{(\ell)}
=\bm\varepsilon_{jk}^{(\ell)}$, and
\begin{equation}
\begin{aligned}
    \bm\Sigma^{(\ell)}&=\gamma_\ell^2\bm B_\ell^{-1}\bm B_\ell^{-\top},\\
    (\bm\Sigma^{(\ell)})^{-1}&=\gamma_\ell^{-2}\bm B_\ell^\top\bm B_\ell,\\
    |\bm\Sigma^{(\ell)}|&=(\gamma_\ell^2)^{n_\ell}.
\end{aligned}
\end{equation}
The unit diagonal makes $\bm B_\ell$ nonsingular, so
$\gamma_\ell^2>0$ ensures that each covariance block is
positive definite. Combining the blocks under the
conditional independence assumption in the mixture model
gives the component log density
\begin{equation}
\begin{aligned}
    \log f_k(\bm y_j;\bm\mu_k,\bm\Sigma)
    &=-\frac{N}{2}\log(2\pi)\\
    &\quad-\frac{1}{2}\sum_{\ell=1}^{L}
    \left\{
    n_\ell\log\gamma_\ell^2
    +\frac{\|\bm B_\ell\bm e_{jk}^{(\ell)}\|^2}
    {\gamma_\ell^2}
    \right\}.
\end{aligned}
\label{eq:component-log-density}
\end{equation}
The quadratic term is the sum of squared innovations,
scaled by their variance, and can be evaluated directly
from the residual recursion.

We consider four covariance specifications that differ
in the sharing of $\phi_\ell$ and $\gamma_\ell^2$ across
conditions (Table~\ref{tab:covariance}). S1 shares both
parameters, S2 shares only the antedependence coefficient,
S3 shares only the innovation variance, and S4 estimates
both parameters separately for each condition.
\begin{table}[t]
    \centering
    \caption{Covariance parameter sharing across conditions.
    Parameters are shared by all clusters under every specification.}
    \label{tab:covariance}
    \setlength{\tabcolsep}{3pt}
    \begin{tabular}{@{}cccc@{}}
        \toprule
        Model & \shortstack{Antedependence\\coefficient} & \shortstack{Innovation\\variance}
        & Parameters \\
        \midrule
        S1 & $\phi_\ell=\phi$ & $\gamma_\ell^2=\gamma^2$ & $2$ \\
        S2 & $\phi_\ell=\phi$ & $\gamma_\ell^2$ & $L+1$ \\
        S3 & $\phi_\ell$ & $\gamma_\ell^2=\gamma^2$ & $L+1$ \\
        S4 & $\phi_\ell$ & $\gamma_\ell^2$ & $2L$ \\
        \botrule
    \end{tabular}
\end{table}
The corresponding covariance parameter sets are
\begin{equation}
\begin{aligned}
    \bm\Theta_\Sigma^{(S1)}
    &=\{\phi,\gamma^2\},\\
    \bm\Theta_\Sigma^{(S2)}
    &=\{\phi,\gamma_1^2,\ldots,\gamma_L^2\},\\
    \bm\Theta_\Sigma^{(S3)}
    &=\{\phi_1,\ldots,\phi_L,\gamma^2\},\\
    \bm\Theta_\Sigma^{(S4)}
    &=\{\phi_1,\ldots,\phi_L,
    \gamma_1^2,\ldots,\gamma_L^2\}.
\end{aligned}
\end{equation}
All four specifications retain the block-diagonal
covariance structure. Sharing parameters across conditions
therefore constrains their within-condition dependence
without introducing cross-condition residual covariance.

\subsection{EM algorithm}

For a fixed number of clusters $K$ and covariance
specification, we estimate the parameters by an
expectation--maximization (EM) procedure
\citep{DempsterEtAl1977}. The unobserved
quantity is the cluster membership $C_j$ of each feature.
The E-step computes its posterior membership probabilities;
the M-step updates the mixture proportions, mean profiles
and covariance parameters using these probabilities as
weights. Because the mean and covariance parameters are
coupled, the M-step is implemented by alternating
conditional updates, following the expectation/conditional
maximization framework \citep{MengRubin1993}.

\paragraph{Initialization.}
We first apply $K$-means to the concatenated feature
profiles $\bm y_1,\ldots,\bm y_p$ to obtain provisional
clusters. The initial mixture proportions are their
relative sizes, and the initial mean coefficients are
obtained by ordinary least squares as described in
Appendix~\ref{app:ols}. The initial covariance parameters
are then computed from the corresponding residuals under
the chosen S1--S4 specification.

\paragraph{E-step.}
Let $\bm\Theta^{(t)}$ denote the parameter estimates at
iteration $t$. The posterior probability that feature $j$
belongs to cluster $k$ is
\begin{equation}
    w_{jk}^{(t)}
    =\Pr(C_j=k\mid\bm y_j;\bm\Theta^{(t)})
    =\frac{
    \pi_k^{(t)}f_k(\bm y_j;\bm\mu_k^{(t)},\bm\Sigma^{(t)})
    }{
    \sum_{h=1}^{K}\pi_h^{(t)}
    f_h(\bm y_j;\bm\mu_h^{(t)},\bm\Sigma^{(t)})
    }.
    \label{eq:responsibilities}
\end{equation}
These probabilities satisfy $\sum_{k=1}^{K}w_{jk}^{(t)}=1$.
Their sum over features,
$N_k^{(t)}=\sum_{j=1}^{p}w_{jk}^{(t)}$, is the effective
size of cluster $k$. The component log densities are
evaluated using equation~\eqref{eq:component-log-density},
and the denominator is evaluated by the log-sum-exp
identity for numerical stability.

\paragraph{M-step.}
With $w_{jk}^{(t)}$ held fixed, the expected
complete-data log likelihood, up to an additive constant,
is
\begin{equation}
\begin{aligned}
    Q(\bm\Theta\mid\bm\Theta^{(t)})
    ={}&\sum_{k=1}^{K}N_k^{(t)}\log\pi_k\\
    &-\frac{1}{2}\sum_{\ell=1}^{L}
    \left\{
    pn_\ell\log\gamma_\ell^2
    +\frac{1}{\gamma_\ell^2}
    \sum_{j=1}^{p}\sum_{k=1}^{K}w_{jk}^{(t)}
    \|\bm B_\ell\bm e_{jk}^{(\ell)}\|^2
    \right\}.
\end{aligned}
\label{eq:em-objective}
\end{equation}
Here, $\bm e_{jk}^{(\ell)}$ and $\bm B_\ell$ depend on
the candidate mean and antedependence parameters,
respectively. The mixture proportions have the update
\begin{equation}
    \pi_k^{(t+1)}=\frac{N_k^{(t)}}{p},
    \qquad k=1,\ldots,K.
\end{equation}
We update the remaining parameters in the following order.
Throughout these conditional updates, the posterior
weights remain fixed at their E-step values.

\emph{Mean profiles.}
For each cluster and condition, form the weighted mean
profile
\begin{equation}
    \overline{\bm y}_k^{(\ell,t)}
    =\frac{1}{N_k^{(t)}}
    \sum_{j=1}^{p}w_{jk}^{(t)}\bm y_j^{(\ell)}.
\end{equation}
Let $\bm X_\ell=(\bm 1_{n_\ell},\bm z_\ell)$, where
$\bm z_\ell=(z_1^{(\ell)},\ldots,z_{n_\ell}^{(\ell)})^\top$,
and let
$\bm\theta_k^{(\ell)}=(\alpha_k^{(\ell)},
\beta_k^{(\ell)})^\top$.
For fixed covariance parameters, maximizing
$Q$ with respect to this coefficient vector is equivalent
to minimizing
$\|\bm B_\ell(\overline{\bm y}_k^{(\ell,t)}
-\bm X_\ell\bm\theta_k^{(\ell)})\|^2$.
Thus the unconstrained update is
\begin{equation}
    \bm\theta_k^{(\ell),\mathrm{new}}
    =\bigl(\bm X_\ell^\top\bm B_\ell^\top
    \bm B_\ell\bm X_\ell\bigr)^{-1}
    \bm X_\ell^\top\bm B_\ell^\top\bm B_\ell
    \overline{\bm y}_k^{(\ell,t)}.
    \label{eq:mean-update}
\end{equation}
This is a generalized least-squares fit with only two
coefficients per cluster and condition. The implementation
adds a small diagonal regularizer to the normal matrix.
If a slope bound is active, the slope is restricted to
its allowed interval and the intercept is refitted with
that slope held fixed.

\emph{Covariance parameters.}
Using the updated mean profiles, recompute the residuals
$e_{jk}^{(\ell)}(i)$. Define three weighted residual sums
for each condition:
\begin{equation}
\begin{aligned}
    U_\ell
    &=\sum_{j=1}^{p}\sum_{k=1}^{K}w_{jk}^{(t)}
    \sum_{i=2}^{n_\ell}[e_{jk}^{(\ell)}(i-1)]^2,\\
    V_\ell
    &=\sum_{j=1}^{p}\sum_{k=1}^{K}w_{jk}^{(t)}
    \sum_{i=2}^{n_\ell}
    e_{jk}^{(\ell)}(i)e_{jk}^{(\ell)}(i-1),\\
    W_\ell
    &=\sum_{j=1}^{p}\sum_{k=1}^{K}w_{jk}^{(t)}
    \sum_{i=1}^{n_\ell}[e_{jk}^{(\ell)}(i)]^2.
\end{aligned}
\end{equation}
The weighted sum of squared innovations is then
\begin{equation}
    R_\ell(\phi_\ell)
    =\sum_{j=1}^{p}\sum_{k=1}^{K}w_{jk}^{(t)}
    \|\bm B_\ell\bm e_{jk}^{(\ell)}\|^2
    =W_\ell-2\phi_\ell V_\ell+\phi_\ell^2U_\ell.
\end{equation}
For condition-specific antedependence coefficients
(S3 and S4), their conditional updates are
\begin{equation}
    \phi_\ell^{\mathrm{new}}=\frac{V_\ell}{U_\ell}.
\end{equation}
For a shared coefficient (S1 and S2), the update pools
conditions with weights given by their current innovation
precisions:
\begin{equation}
    \phi^{\mathrm{new}}
    =\frac{\sum_{\ell=1}^{L}V_\ell/\gamma_\ell^2}
    {\sum_{\ell=1}^{L}U_\ell/\gamma_\ell^2}.
    \label{eq:shared-phi-update}
\end{equation}
Under S1, the variances are equal and this reduces to
$\sum_\ell V_\ell/\sum_\ell U_\ell$.
Under S2, the condition-specific variance weights are
retained. The updated coefficient is restricted to a
closed interval inside $(-1,1)$.

Given the updated antedependence coefficients, the
innovation variances are updated as
\begin{equation}
\begin{aligned}
    (\gamma_\ell^2)^{\mathrm{new}}
    &=\frac{R_\ell(\phi_\ell^{\mathrm{new}})}{pn_\ell},
    &&\text{for S2 and S4},\\
    (\gamma^2)^{\mathrm{new}}
    &=\frac{\sum_{\ell=1}^{L}
    R_\ell(\phi_\ell^{\mathrm{new}})}{pN},
    &&\text{for S1 and S3}.
\end{aligned}
\label{eq:variance-update}
\end{equation}
The denominators count the innovations contributing to
each variance estimate: $pn_\ell$ for a single condition
and $pN$ when all conditions share a variance.
A positive lower bound is imposed on the variances to
maintain numerical stability. Mean and covariance updates
are alternated within the M-step, using the most recent
parameter values, until the inner stopping criterion or
iteration limit is reached.

\paragraph{Convergence and cluster assignment.}
After each M-step, we recompute the posterior weights and
the observed-data log likelihood
\begin{equation}
    \mathcal L^{(t)}
    =\sum_{j=1}^{p}\log\left\{
    \sum_{k=1}^{K}\pi_k^{(t)}
    f_k(\bm y_j;\bm\mu_k^{(t)},\bm\Sigma^{(t)})
    \right\}.
\end{equation}
The outer iteration stops when
$|\mathcal L^{(t+1)}-\mathcal L^{(t)}|/p\leq\epsilon$
or the maximum number of iterations is reached.
Within the M-step, the inner criterion is the largest
absolute change in the mean coefficients, antedependence
coefficients and innovation standard deviations.
The default implementation uses $\epsilon=10^{-3}$,
at most 200 outer iterations and two inner update cycles.
When multiple initializations are requested, the fit with
the largest observed-data log likelihood is retained.
The final assignment is
\begin{equation}
    \widehat C_j=\operatorname*{arg\,max}_{1\leq k\leq K}
    \widehat w_{jk},
\end{equation}
where $\widehat w_{jk}$ is the posterior membership
probability evaluated at the retained parameter estimates.
The full posterior probability vector is retained to
quantify uncertainty in each assignment.

The fitting procedure for a fixed $(K,S)$ is summarized
in Algorithm~\ref{alg:fast-funclu}.
\begin{algorithm}[t]
\caption{Fast FunClu for fixed $K$ and $S$}
\label{alg:fast-funclu}
\begin{algorithmic}[1]
\Require Profiles $\bm Y$, coordinates $\{\bm z_\ell\}$, $K$,
covariance specification $S$, number of starts $M$,
tolerance $\epsilon$, iteration limits $T$ and $H$.
\Ensure Fitted parameters $\widehat{\bm\Theta}$, posterior
weights $\widehat w_{jk}$ and module labels $\widehat C_j$.
\State Cache $y_j^{(\ell)}(i)^2$ and
$y_j^{(\ell)}(i)y_j^{(\ell)}(i-1)$.
\For{$m=1,\ldots,M$}
    \State Initialize labels by $K$-means; initialize means
    by OLS and covariance parameters under $S$.
    \State Evaluate the innovation-form likelihood and
    posterior weights $w_{jk}$.
    \For{$t=1,\ldots,T$}
        \State Store $\mathcal L_{\mathrm{old}}\gets\mathcal L$.
        \State Compute $N_k=\sum_j w_{jk}$,
        $\pi_k\gets N_k/p$ and weighted mean profiles.
        \For{$h=1,\ldots,H$}
            \State Store the current mean and covariance parameters.
            \State Update mean coefficients by GLS
            using equation~\eqref{eq:mean-update}.
            \State Compute weighted residual statistics
            $U_\ell,V_\ell,W_\ell$ using cached moments.
            \If{$S\in\{S3,S4\}$}
                \State $\phi_\ell\gets V_\ell/U_\ell$
                for each condition $\ell$.
            \Else
                \State Update shared $\phi$ by
                equation~\eqref{eq:shared-phi-update}.
            \EndIf
            \State Restrict coefficients to their allowed interval.
            \State Update innovation variances by
            equation~\eqref{eq:variance-update}; enforce positivity.
            \If{maximum parameter change $\leq\epsilon$}
                \State \textbf{break}
            \EndIf
        \EndFor
        \State Re-evaluate $\mathcal L$ and $w_{jk}$
        using innovation-form densities and log-sum-exp.
        \If{$|\mathcal L-\mathcal L_{\mathrm{old}}|/p\leq\epsilon$}
            \State \textbf{break}
        \EndIf
    \EndFor
    \State Retain this start if its final likelihood is largest.
\EndFor
\State \Return retained $\widehat{\bm\Theta}$ and $\widehat w_{jk}$,
with $\widehat C_j=\arg\max_k\widehat w_{jk}$.
\end{algorithmic}
\end{algorithm}

\subsection{Model selection}

The number of clusters and the degree of covariance
parameter sharing are selected jointly. Let
$\mathcal K$ denote the candidate set of cluster numbers
and $\mathcal S=\{S1,S2,S3,S4\}$ the covariance
specifications in Table~\ref{tab:covariance}.
For each pair $(K,S)\in\mathcal K\times\mathcal S$,
we fit the mixture model using the EM procedure above
and denote the fitted observed-data log likelihood by
$\widehat{\mathcal L}_{K,S}
=\log L(\widehat{\bm\Theta}_{K,S}\mid\bm Y)$.
All candidates are fitted to the same feature profiles
and niche coordinates.

The total number of free parameters is
\begin{equation}
\begin{gathered}
    q_{K,S}=(K-1)+2KL+q_\Sigma(S),\\
    q_\Sigma(S)=
    \begin{cases}
        2, & S=S1,\\
        L+1, & S=S2\text{ or }S3,\\
        2L, & S=S4.
    \end{cases}
\end{gathered}
    \label{eq:parameter-count}
\end{equation}
The three terms count the independent mixture
proportions, the two mean coefficients for each cluster
and condition, and the covariance parameters,
respectively. The covariance contribution is counted
once because these parameters are shared by all clusters.

We select the model by minimizing the Bayesian
information criterion (BIC) \citep{Schwarz1978},
\begin{equation}
    \operatorname{BIC}(K,S)
    =-2\widehat{\mathcal L}_{K,S}+q_{K,S}\log p.
    \label{eq:bic}
\end{equation}
The sample size in the penalty is $p$, the number of
feature profiles contributing to the mixture likelihood.
Each profile is an $N$-dimensional observation whose
within-profile dependence is represented by
$\bm\Sigma$. The selected pair is
\begin{equation}
    (\widehat K,\widehat S)
    =\operatorname*{arg\,min}_{(K,S)\in\mathcal K\times\mathcal S}
    \operatorname{BIC}(K,S).
\end{equation}
This joint comparison balances the fit gained by
additional clusters or condition-specific covariance
parameters against their parameter cost, following the
model-based clustering strategy of \citet{FraleyRaftery2002}.
If either $K$ or $S$ is specified in advance, the
comparison is restricted to the remaining candidates.

For each candidate, we also report the Akaike
information criterion \citep{Akaike1974},
\begin{equation}
    \operatorname{AIC}(K,S)
    =-2\widehat{\mathcal L}_{K,S}+2q_{K,S},
\end{equation}
and its BIC difference from the selected model,
$\Delta\operatorname{BIC}(K,S)
=\operatorname{BIC}(K,S)
-\operatorname{BIC}(\widehat K,\widehat S)$.
The comparison table includes convergence status and the
smallest assigned cluster size for each fit.
BIC determines the final model; AIC and the fit diagnostics
are retained to assess the competing solutions.

\section{Simulation experiments}
\label{sec:simulations}

\subsection{Computational gains from closed-form SAD(1) updates}
\label{sec:computational-efficiency}

Closed-form SAD(1) updates substantially accelerated
covariance estimation and reduced complete fitting time.
We quantified these gains through matched-model fitting,
conditional covariance solving and original-software
comparisons.
We first compared implementations of the same mixture
model, using identical initial parameter values, GLS mean
updates, parameter constraints and stopping criteria.
The proposed implementation was compared with blockwise
numerical covariance optimization, joint numerical
covariance optimization, and a version that evaluated
the closed updates directly from the original curves
instead of cached quadratic statistics. The numerical
comparators used analytic gradients and parameter scaling.
The reference configuration had $L=3$ conditions,
$p=600$ features, $n_\ell=40$ samples per condition and
$K=6$ clusters. We varied $p$, $n_\ell$ or $K$ separately
and added correctly specified S1--S3 configurations,
giving nine configurations with 30 independent datasets
each. The generating parameters and computational
settings are given in Appendix~\ref{app:computational-benchmarks}.

\begin{figure*}[hbp]
    \centering
    \includegraphics[width=\linewidth]{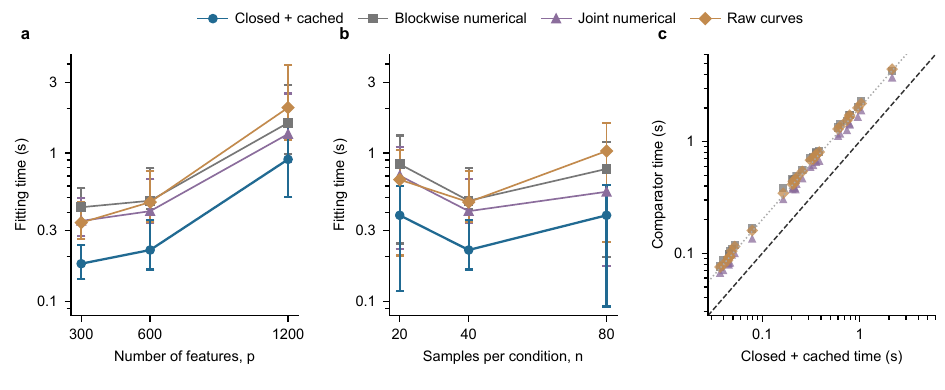}
    \caption{Computational cost under the same mixture
    model. (a,b) Median fitting time versus feature count
    $p$ and samples per condition $n$, respectively, with
    95\% dataset-level bootstrap intervals (5,000 resamples).
    Each setting has 30 independent datasets and three
    common numerical starts per method. The reference
    has $L=3$, $p=600$, $n_\ell=40$, $K=6$ and S4
    covariance; only the indicated dimension changes.
    All methods share constraints and stopping criteria.
    Timing includes preparation and all starts, excluding
    common initialization. (c) Paired
    fitting times at the reference setting, with 30 points
    per comparator. The dashed and dotted lines denote
    $y=x$ and $y=2x$. Vertical axes in (a,b) and both
    axes in (c) are logarithmic. Reference datasets recur
    across panels and do not constitute extra replicates.}
    \label{fig:matched-computation}
\end{figure*}

Closed updates reduced complete fitting time by
1.87--2.37-fold relative to blockwise numerical
optimization across the nine configurations
(Figs.~\ref{fig:matched-computation} and
\ref{fig:matched-speed-distributions}). In the reference
configuration, the median paired speedup was 2.19-fold
(95\% bootstrap confidence interval, 2.15--2.21).
Speedups relative to joint numerical optimization were
1.60--1.98-fold. All 1,080 multi-start fits completed,
and the largest absolute difference in final
log likelihood per feature was $2.69\times10^{-12}$;
cluster assignments were identical between paired
implementations. Including initialization retained a
1.81--2.28-fold speedup over blockwise numerical
optimization. The accelerated fits retained agreement
in likelihood and identical cluster assignments.

Median fitting time increased with the number of features
over the tested range (Fig.~\ref{fig:matched-computation}a),
whereas its variation with samples per condition was
non-monotonic (Fig.~\ref{fig:matched-computation}b).
At the reference configuration,
paired times placed every comparator above the equal-time
line (Fig.~\ref{fig:matched-computation}c). The paired
ratio distributions also remained above one across the
size configurations and correctly specified S1--S4
constraints (Fig.~\ref{fig:matched-speed-distributions}),
showing that the computational gain persisted across
datasets and covariance specifications.

We next examined covariance solving within a single R
environment, using the conditional objectives from the
\texttt{idopNetwork} FunClu implementation. Six
configurations combined linear or power-law generating
means with single-condition, two-condition and larger
two-condition tasks. For ten predetermined datasets per
configuration, we fixed the same mean curves and posterior
weights and compared closed updates with high-precision
BFGS optimization of the original conditional objective.
The closed implementation included the cost of constructing
weighted residual statistics from the curves. Across these
60 targets, median speedups ranged from 91.5 to 201.4-fold
(Appendix~\ref{app:computational-benchmarks},
Fig.~\ref{fig:same-r-covariance}b). Paired timings
lay around or above the hundredfold reference line
(Fig.~\ref{fig:same-r-covariance}a). The largest conditional
objective difference per feature was $1.31\times10^{-8}$,
below the predefined $10^{-6}$ threshold in every case.
The closed-form solver used weighted residual statistics
and innovation-form evaluation to replace repeated
numerical optimization and dense Gaussian density
calculations with direct parameter updates.

\begin{figure*}[hbp]
    \centering
    \includegraphics[width=\linewidth]{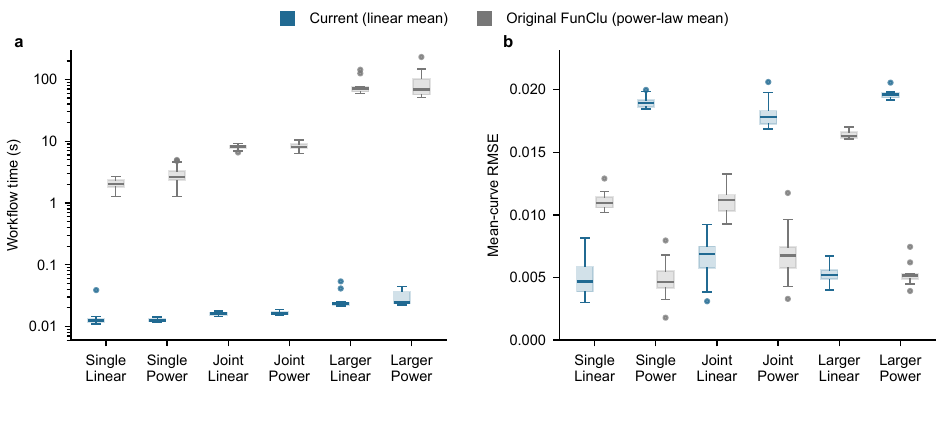}
    \caption{Workflow time and mean recovery relative
    to the upstream FunClu software. (a) Complete
    three-start workflow time, including common $K$-means
    and model-specific initialization, on a logarithmic
    axis. Process startup, source loading and input
    reading are excluded. (b) Label-matched mean-curve
    RMSE over clusters, conditions and sample positions.
    Single and Joint tasks use $(p,n_\ell,K)=(300,20,3)$
    with one and two conditions, respectively; Larger
    uses $(600,40,6)$ with two conditions. Linear and
    Power denote the generating mean family. There are
    30 independent datasets per baseline configuration
    and ten per larger configuration. Boxes span the
    quartiles, center lines mark medians, whiskers reach
    the most extreme observations within 1.5 interquartile
    ranges, and circles show all observations beyond the
    whiskers. Input responses and label starts are shared;
    mean families, languages, parameter domains and
    stopping rules differ between workflows.}
    \label{fig:upstream-quality}
\end{figure*}

Finally, we compared the current implementation with
the upstream functions \texttt{fun\_clu} and
\texttt{bifun\_clu} in \texttt{idopNetwork}
version~0.1.2. The source functions were unmodified. The
single- and two-condition tasks each used 300 features,
20 samples per condition and three clusters; the larger
two-condition task used 600 features, 40 samples per
condition and six clusters. Both linear and power-law
generating means were considered, with 30 independent
datasets per baseline configuration and ten per larger
configuration, giving 140 paired datasets. Each workflow
received three common data-derived label starts and
model-specific mean and covariance initializations.
All workflows completed. Including initialization, the
median paired speedups were 156--211-fold for single
conditions, 488--499-fold for two conditions, and
2,811--2,913-fold for the larger tasks
(Fig.~\ref{fig:upstream-quality}a;
Table~\ref{tab:upstream-workflows}). The upstream software
fitted power-law means, whereas the current implementation
fitted linear means on the modeled response scale.

Both workflows achieved an adjusted Rand index
(ARI; \citealp{HubertArabie1985}) of 1 in every dataset
in this last experiment. Mean-curve
errors were smaller for the current model under linear
generation and for the upstream model under power-law
generation (Fig.~\ref{fig:upstream-quality}b).

\subsection{Module recovery relative to \texorpdfstring{$K$}{K}-means and an unrestricted GMM}
\label{sec:clustering-quality}

FunClu improved module recovery over $K$-means and a
Gaussian mixture with unrestricted mean vectors and a shared
full covariance matrix when the generating mean structure
matched the model. We compared these methods with an
independent-residual FunClu variant, which retained the same
mean model but fixed all SAD(1) coefficients to zero.
The comparison used three conditions, 20 ordered positions
per condition and four balanced modules. We varied the
number of training curves, $p\in\{300,600,1200\}$, and the
residual coefficient, $\phi\in\{0,0.4,0.8\}$, with 30
independent datasets per setting. Mean separation and
average marginal noise variance were held constant.
All methods received the same response matrix, five common
initial partitions and the true $K=4$. Recovery was evaluated
using ARI on 2,000 independently generated test curves per
dataset and label-matched mean-curve RMSE. This experiment
assessed recovery at a given $K$; generating parameters and
fitting details appear in Appendix~\ref{app:clustering-quality}.

\begin{figure*}[!htbp]
    \centering
    \includegraphics[width=0.92\linewidth]{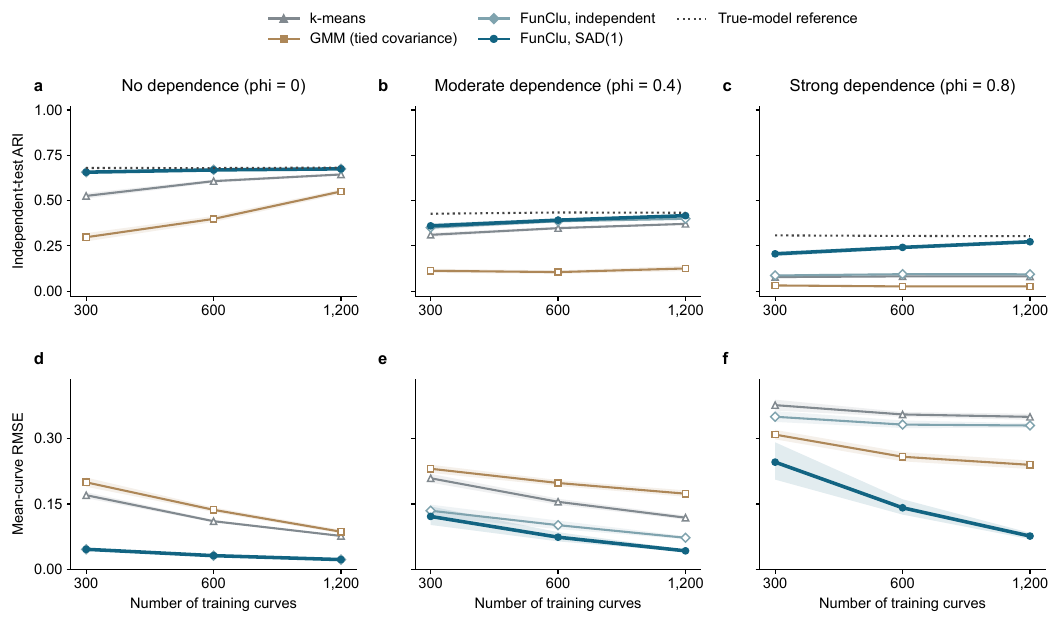}
    \caption{Module and mean-curve recovery under linear
    generating means. Top: ARI on independently generated
    test curves; bottom: label-matched mean-curve RMSE.
    Columns indicate the generating SAD(1) coefficient.
    Each point summarizes 30 independent training datasets,
    with 2,000 independent test curves per dataset.
    Shaded bands are pointwise 95\% bootstrap intervals for
    the mean, based on 5,000 dataset-level resamples.
    All methods use $K=4$ and five shared initial partitions.
    GMM denotes unrestricted mean vectors with a shared
    full covariance matrix. The independent FunClu variant
    fixes $\phi=0$. Dotted lines show classification using
    the true generating parameters as a difficulty reference.}
    \label{fig:linear-recovery}
\end{figure*}

FunClu achieved higher mean test ARI and lower mean-curve
RMSE than both baselines throughout the linear-mean grid
(Fig.~\ref{fig:linear-recovery}). At $p=600$ and
$\phi=0.8$, mean test ARI was 0.241 for FunClu,
compared with 0.082 for $K$-means, 0.026 for the full
covariance GMM and 0.094 for independent-residual FunClu.
The paired ARI gain over $K$-means was 0.160
(95\% dataset-level bootstrap interval, 0.150--0.170),
and that over independent-residual FunClu was 0.148
(0.137--0.159). Mean-curve RMSE was 0.142, compared
with 0.355, 0.258 and 0.332, respectively.
The two FunClu variants performed similarly without
residual correlation, whereas SAD(1) provided a clear
additional benefit under strong correlation. Increasing
the number of training curves reduced FunClu's mean-curve
error at each correlation level.

The benefit depended on the mean structure. In a matched
separation control with quadratic mean components,
$K$-means achieved higher test ARI than FunClu at
$\phi=0$ and $0.4$; at $\phi=0$, the corresponding
values were 0.640 and 0.378. Under strong correlation,
FunClu retained a classification advantage but had larger
mean-curve error (Fig.~\ref{fig:curved-mean-control}).
Thus, the structured means improved recovery when
appropriate, while SAD(1) contributed most clearly when
residual dependence was strong. These comparisons
complement the computational benchmarks by identifying
the data structures for which the fitted model was effective.

\section{Real Data Applications}
\label{sec:real-data}

We examined single-, two- and three-condition datasets in turn:
spatial gut microbiota, paired leaf--root microbiota and
gut multi-omics profiles measured at three visits.

\subsection{Gut microbiota benchmark}
\label{sec:gut-benchmark}

We first analysed the gut microbiota example distributed
with idopNetwork \citep{DongEtAl2023}, derived from the
spatial microbiota study of \citet{CaoEtAl2022}.
From the package's saved analysis objects, we recovered
the 48-feature input used in its precomputed gut example,
covering 20 samples and including the aggregate ``Others''
category. These features had nonzero abundance in at least
15 of the 20 samples.
For this abundance dataset, we modeled
$y=\log_{10}(1+\text{abundance})$ without feature-wise
standardization, retaining between-feature abundance
levels. The niche index was one plus total retained abundance
per sample, matching the saved example's clustering coordinate.
We compared $K=1,\ldots,20$ with five initializations per
candidate and required every module to contain at least
two maximum-posterior assigned features. Among fitted
candidates satisfying this requirement, BIC selected seven
modules (BIC $=2,198.39$); the six-module fit was higher by
7.45. Without the size requirement, BIC selected eight
modules, with a BIC 6.47 lower but two singleton modules.

\begin{figure*}[!htbp]
    \centering
    \includegraphics[width=\linewidth]{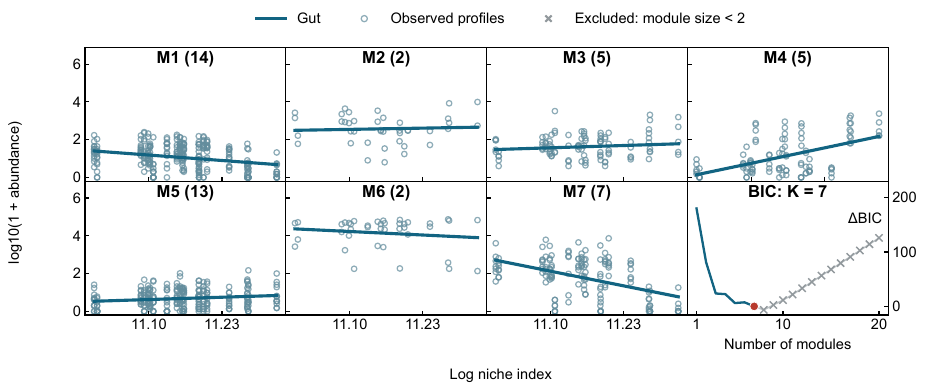}
    \caption{Abundance-preserving analysis of the
    idopNetwork gut microbiota example. Dark solid lines
    show the seven selected component profiles; pale hollow
    circles show all 960 observed feature-by-sample values
    from the 48 retained features and 20 samples.
    Parentheses give feature counts. Values are shown
    directly on the modeled $y=\log_{10}(1+x)$ scale,
    without back-transformation. All module panels share
    the same linear vertical axis limits and tick locations.
    M6 includes the aggregate ``Others'' profile.
    The BIC panel compares $K=1,\ldots,20$; gray crosses
    mark fitted candidates with any module containing fewer
    than two assigned features. The red marker identifies
    $K=7$, the BIC minimum among admissible candidates.
    Differences are measured from this selected fit;
    an excluded candidate can therefore have a negative
    $\Delta\mathrm{BIC}$. Circles show observed member
    variation, rather than confidence intervals.}
    \label{fig:idop-gut}
\end{figure*}

The seven modules contained 14, 2, 5, 5, 13, 2 and 7
features, respectively, with no singleton assignments
(Fig.~\ref{fig:idop-gut}). Four centers increased and
three decreased along the niche coordinate. M4 had the
steepest increasing standardized-coordinate slope
($\beta=0.515$), and M7 the steepest decreasing slope
($\beta=-0.491$). M6 grouped the aggregate ``Others''
feature with \textit{Escherichia coli}; it therefore
includes a pooled abundance category rather than two
individual taxa. The mean maximum posterior membership
probability was 0.916. The candidate-selection rule and
preprocessing sensitivity analyses are detailed in
Appendix~\ref{app:idop-data}.

\subsection{Joint leaf and root microbiota analysis}
\label{sec:plant-benchmark}

To examine joint clustering across organs, we analysed
 the idopNetwork mustard microbiota example, originating
from the study of \citet{WagnerEtAl2016} on
\textit{Boechera stricta}. The distributed data contained
3,263 operational taxonomic units (OTUs) and 88 matched
leaf--root sample pairs. The package's filtering rule
retained 3,137 OTUs, each represented by a joint profile
of length 176. Within each organ, we standardized each
feature, applied a global shift to minimum one and
modeled the logarithm of the shifted values. We fitted S4 with
organ-specific covariance parameters and a common OTU
assignment. With five initializations per candidate,
we compared $K=1,\ldots,40$.
BIC selected eleven modules (BIC $=163,444.46$);
 the ten-module solution was higher by 15.69.

\begin{figure*}[!htbp]
    \centering
    \includegraphics[width=\linewidth]{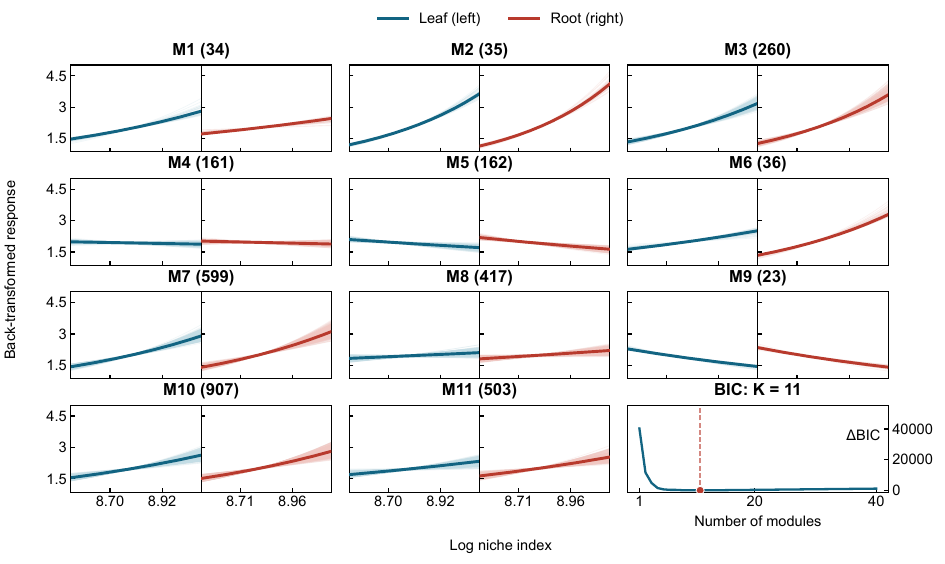}
    \caption{Joint leaf and root microbiota analysis after
    within-organ feature standardization and positive shifting.
    Each module is displayed as an adjacent pair of panels,
    with leaves on the left and roots on the right.
    A common module heading and member count appear above
    each pair. Panels within a pair touch; separate modules
    are spaced apart. Blue and red solid lines show
    leaf and root component profiles, respectively;
    pale lines show every assigned OTU's fitted
    power function. Each organ contributes 88 samples.
    Parentheses give OTU counts, totalling 3,137.
    Curves span each organ's observed log niche range.
    The vertical axis is the back-transformed modeled
    response, on the shifted standardized scale.
    All module panels share the same vertical limits and
    ticks, including both organs. The final panel
    shows the fixed-S4 $\Delta\mathrm{BIC}$ scan over $K=1,\ldots,40$,
    selecting $K=11$. Pale curves show member heterogeneity,
    rather than inferential uncertainty.}
    \label{fig:idop-plant-1}
\end{figure*}

The eleven-module fit identified shared profile directions
with organ-specific slopes (Fig.~\ref{fig:idop-plant-1}).
Eight module centers increased in both organs and three
 decreased in both. M6, containing 36 OTUs, showed a
steeper standardized slope in roots than in leaves
($\beta_{\mathrm{root}}=0.243$ versus
$\beta_{\mathrm{leaf}}=0.125$), whereas M1 showed the
reverse pattern (0.098 versus 0.184).
M9, containing 23 OTUs, had the strongest decreasing
center in both organs. The three largest modules,
M10, M7 and M11, contained 907, 599 and 503 OTUs,
respectively, accounting for 64.0\% of the retained OTUs.
The mean maximum posterior membership probability was
0.804. The joint model therefore retained a common
module partition while describing distinct niche-associated
profiles in leaves and roots.

\subsection{Three-condition gut multi-omics analysis}
\label{sec:human-multiomics}

\paragraph{Study population and multi-omics data.}

We applied the proposed model to gut multi-omics data
from a randomized, double-blind, placebo-controlled study
of collegiate endurance athletes \citep{ZhangEtAl2026}.
The source study enrolled 150 participants and evaluated
a combined probiotics--yeast extract intervention (CPYE)
over a six-week phase, with assessments at baseline,
Week~3 and Week~6. Our analysis used the 149 participants
who completed this phase: 99 in the CPYE group and 50
in the placebo group, comprising 66 men and 83 women.
The same participants were represented at all three visits.
The analysis tables were assembled from the published
short-term multi-omics source data, available at
\url{https://doi.org/10.6084/m9.figshare.32043930}
\citep{Zhang2026SourceData}.

The dataset contained 3,796 features: 199 bacterial taxa,
199 fungal taxa, 30 viral features, 2,098 microbial
functional features and 1,270 metabolite features.
All features were available at each visit, giving three
$149\times3,796$ matrices with no missing entries.
These matrices retained all feature columns in the five
source tables. We treated the three visits as conditions
and constructed a joint profile of length $N=447$ for each
feature by concatenating its visit-specific niche-ordered
profiles. Clustering was performed jointly across all
participants, with a common feature assignment across
visits. Treatment and sex labels were not included in
this pooled clustering analysis.

\paragraph{Model fitting and selection.}

\begin{figure*}[!htbp]
    \centering
    \includegraphics[width=\linewidth]{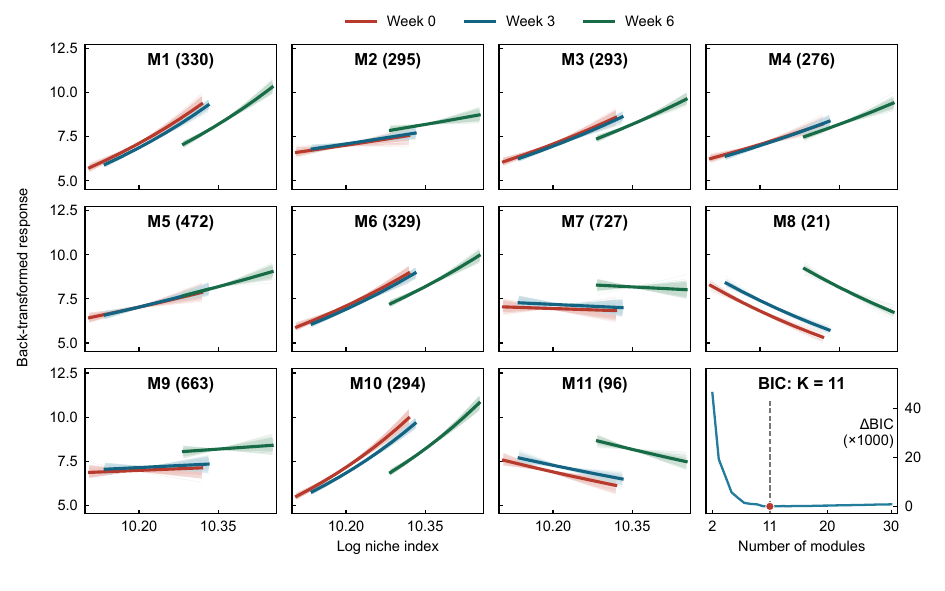}
    \caption{Niche-associated profiles of the eleven selected
    modules. Dark lines show the back-transformed component
    profiles at Week~0, Week~3 and Week~6, shown as solid
    lines in red, blue and green, respectively. Parentheses give feature counts
    (3,796 in total). Curves span each visit's observed log niche
    range, with 149 participants per visit. Pale lines show
    every assigned member's fitted power function, sampled at
    160 grid points. Member coefficients are obtained by
    ordinary least squares on the modeled response and
    standardized niche coordinate; both member and component
    curves are then exponentiated for display. The vertical
    axis is therefore the back-transformed response, not raw
    abundance. Module panels share
    axis scales. The final panel shows the BIC difference from
    the minimum over $K=2,\ldots,30$ under S4; the marker identifies
    $K=11$. Profiles describe within-visit niche associations,
    rather than temporal trajectories or tests of visit effects.}
    \label{fig:real-data-modules}
\end{figure*}

For this application, we used the S4 covariance
specification, allowing the antedependence coefficient
and innovation variance to differ among visits.
We compared $K=2,\ldots,30$ clusters using BIC.
The minimum occurred at $K=11$, with
$\operatorname{BIC}=-2{,}817{,}008.40$ and 82 free
parameters. The adjacent ten- and twelve-cluster models
had BIC values higher by 43.58 and 41.12, respectively.
The selected solution therefore retained eleven modules,
denoted M1--M11 (Fig.~\ref{fig:real-data-modules}).

With $K$-means initialization, the selected fit met the
convergence criterion after 92 EM iterations and attained
an observed-data log likelihood of $1{,}408{,}842.11$.
The average and median maximum posterior membership
probabilities were 0.853 and 0.904, respectively;
1,926 features (50.7\%) had an assignment probability of
at least 0.90. Module sizes ranged from 21 to 727
features (Table~\ref{tab:real-modules}).

\paragraph{Module composition and niche-associated profiles.}

The eleven modules separated features with different
niche-associated profiles and different omics
compositions. M7 and M9 were the largest, containing
727 and 663 features, respectively. They included 642
and 504 metabolite features, together accounting for
90.2\% of the metabolite features in the dataset.
Microbial functional features predominated in M1, M3,
M4, M5, M6 and M10. In particular, all 294 members of
M10 were functional features, whereas M1 contained
329 functional features and one fungal taxon.
M8 was the smallest module, with 21 members, of which
17 were fungal taxa. M2 had a more mixed composition,
including 52 bacterial taxa, 42 fungal taxa, nine viral
features, 133 functional features and 59 metabolite
features.

The modules showed increasing, decreasing and nearly flat
profiles along the niche coordinate
(Fig.~\ref{fig:real-data-modules}). The fitted slope signs
were consistent across the three visits.
Eight modules (M1--M6, M9 and M10) had positive
slopes, and three (M7, M8 and M11) had negative slopes.
M10 showed the largest positive standardized niche slopes,
with $\beta=0.111$, $0.114$ and $0.095$ at baseline,
Week~3 and Week~6, respectively. M8 showed the largest
negative slopes, with corresponding values of $-0.083$,
$-0.084$ and $-0.065$. M7 and M9 had comparatively flat
profiles: their fitted slopes ranged from $-0.0085$ to
$-0.0058$ and from $0.0070$ to $0.0090$, respectively.
These slopes describe variation across the standardized
niche coordinate within each visit. The retained slope
directions identify recurring niche-associated patterns
at the three visits; they do not represent rates of change
with chronological time.

The fitted antedependence coefficients were 0.0124,
0.0205 and 0.0227 at baseline, Week~3 and Week~6,
respectively, indicating weak dependence between adjacent
residuals in the niche ordering. The corresponding
innovation variances were 0.01257, 0.01151 and 0.00928.
The shared modules and visit-specific profiles provide
descriptive targets for biological follow-up; this pooled
fit did not test treatment contrasts, sex differences
or microbial interactions.

\begin{table}[t]
    \centering
    \caption{Feature counts and omics composition of the
    eleven modules in the selected S4 model. Module
    assignments were determined by maximum posterior
    membership probability.}
    \label{tab:real-modules}
    \small
    \setlength{\tabcolsep}{1.6pt}
    \begin{tabular}{@{}lrrrrrr@{}}
        \toprule
        Module & Total & Bacteria & Fungi & Viruses
        & Functions & Metabolites \\
        \midrule
        M1 & 330 & 0 & 1 & 0 & 329 & 0 \\
        M2 & 295 & 52 & 42 & 9 & 133 & 59 \\
        M3 & 293 & 0 & 4 & 1 & 288 & 0 \\
        M4 & 276 & 2 & 13 & 1 & 260 & 0 \\
        M5 & 472 & 17 & 32 & 2 & 414 & 7 \\
        M6 & 329 & 0 & 6 & 1 & 322 & 0 \\
        M7 & 727 & 46 & 23 & 7 & 9 & 642 \\
        M8 & 21 & 1 & 17 & 2 & 1 & 0 \\
        M9 & 663 & 79 & 38 & 5 & 37 & 504 \\
        M10 & 294 & 0 & 0 & 0 & 294 & 0 \\
        M11 & 96 & 2 & 23 & 2 & 11 & 58 \\
        \midrule
        Total & 3,796 & 199 & 199 & 30 & 2,098 & 1,270 \\
        \botrule
    \end{tabular}
\end{table}

\section{Conclusion}
\label{sec:conclusions}

Closed-form SAD(1) estimation provides a substantial
computational gain for functional clustering. By
expressing the covariance updates through weighted
residual statistics and evaluating the likelihood in
innovation form, the proposed procedure replaces repeated
numerical covariance optimization with direct updates.
In the same-R conditional comparisons, this formulation
accelerated covariance solving by 91.5--201.4-fold,
with all 60 targets meeting the objective-agreement
criterion. The gain was therefore accompanied by
agreement in the attained conditional objectives.

These savings also improved complete model fitting.
Under matched models and computational settings,
complete fitting was 1.87--2.37 times as fast as
blockwise numerical optimization, with identical cluster
assignments and close agreement in final likelihoods.
The resulting implementation achieved median complete-workflow
speedups of 156--2,913-fold relative to the original
FunClu software across the benchmark configurations.
Together, these results show that closed-form SAD(1)
updates, combined with efficient likelihood evaluation,
provide a fast computational foundation for joint
functional clustering across conditions.

The recovery experiments also showed the statistical
benefit of modeling curve structure. With correctly
specified means, FunClu recovered modules and mean curves
more accurately than $K$-means and a GMM with shared
unrestricted covariance across the tested settings.
SAD(1) provided an additional recovery gain under strong
residual correlation. Curved-mean controls identified
the complementary requirement: the chosen mean family
must capture the dominant trajectory shapes.

\section{Conflicts of interest}
The authors declare that they have no competing interests.

\section{Funding}
This work was supported by XXX (grant number XXX).

\section{Data availability}
The gut and paired leaf--root microbiota datasets were
obtained from the publicly available \texttt{idopNetwork}
package repository at
\url{https://github.com/cxzdsa2332/idopNetwork}
\citep{DongEtAl2023}. The source version, retained
features and samples, and preprocessing used in the
present analyses are documented in the Methods and
Appendix~\ref{app:idop-data}.
The processed human gut multi-omics data are available
from the source study's figshare archive at
\url{https://doi.org/10.6084/m9.figshare.32043930}
\citep{Zhang2026SourceData}.

The repository link and persistent identifier for the
FunClu implementation, simulation scripts, fitted results
and figure source data generated in this study are XXX.

\section{Author contributions statement}
Yu Wang designed the experiments and wrote the manuscript.
Jiaze Guo contributed to writing parts of the manuscript.
Yao Chen and Tianxiang Hao organized and checked the code.
Rongling Wu provided the ideas for the experiments.

\section{Acknowledgments}
We thank the Beijing Institute of Mathematical Sciences
and Applications (BIMSA) and Renmin University of China
(RUC) for their support.

\clearpage
\raggedbottom
\begin{appendices}

\section{OLS initialization of the mean profiles}
\label{app:ols}

For each component $k$ under condition $\ell$, let
$\mathcal J_k^{(0)}$ denote the set of features
provisionally assigned to component $k$ during
initialization, with $p_k^{(0)}$ denoting the number
of features in this set. The OLS estimators are
obtained by minimizing
\begin{equation}
    Q_k^{(\ell)}(\alpha,\beta)
    =
    \sum_{j\in\mathcal J_k^{(0)}}
    \sum_{i=1}^{n_\ell}
    \left[
    y_j^{(\ell)}(\tau_i^{(\ell)})
    -\alpha-\beta z_i^{(\ell)}
    \right]^2.
\end{equation}

Setting the partial derivatives with respect to
$\alpha$ and $\beta$ to zero yields
\begin{equation}
\begin{aligned}
    \sum_{j\in\mathcal J_k^{(0)}}
    \sum_{i=1}^{n_\ell}
    \left[
    y_j^{(\ell)}(\tau_i^{(\ell)})
    -\widehat{\alpha}_k^{(\ell)}
    -\widehat{\beta}_k^{(\ell)}z_i^{(\ell)}
    \right]&=0,\\
    \sum_{j\in\mathcal J_k^{(0)}}
    \sum_{i=1}^{n_\ell}
    z_i^{(\ell)}
    \left[
    y_j^{(\ell)}(\tau_i^{(\ell)})
    -\widehat{\alpha}_k^{(\ell)}
    -\widehat{\beta}_k^{(\ell)}z_i^{(\ell)}
    \right]&=0.
\end{aligned}
\end{equation}

Since the standardized variables satisfy
$\sum_{i=1}^{n_\ell}z_i^{(\ell)}=0$ and
$\sum_{i=1}^{n_\ell}(z_i^{(\ell)})^2=n_\ell$,
the OLS estimators simplify to
\begin{equation}
\begin{aligned}
    \widehat{\alpha}_k^{(\ell)}
    &=
    \frac{1}{p_k^{(0)}n_\ell}
    \sum_{j\in\mathcal J_k^{(0)}}
    \sum_{i=1}^{n_\ell}
    y_j^{(\ell)}(\tau_i^{(\ell)}),\\
    \widehat{\beta}_k^{(\ell)}
    &=
    \frac{1}{p_k^{(0)}n_\ell}
    \sum_{j\in\mathcal J_k^{(0)}}
    \sum_{i=1}^{n_\ell}
    z_i^{(\ell)}
    y_j^{(\ell)}(\tau_i^{(\ell)}).
\end{aligned}
\end{equation}

The fitted initial mean profile is
\begin{equation}
    \widehat{\mu}_k^{(\ell)}(\tau_i^{(\ell)})
    =\widehat{\alpha}_k^{(\ell)}
    +\widehat{\beta}_k^{(\ell)}z_i^{(\ell)}.
\end{equation}

These OLS estimates are used only to construct
approximate initial mean profiles. The final mean
parameters are estimated by maximizing the mixture
likelihood within the EM algorithm.

\section{Computational experiment details}
\label{app:computational-benchmarks}

\subsection{Common-model comparisons}

Profiles were generated from the linear-mean Gaussian
mixture with independent conditions and finite-start
SAD(1) residuals, as defined in
Eqs.~\eqref{eq:mean-profile} and \eqref{eq:sad-recursion}.
The reference configuration used $L=3$, $p=600$,
$n_\ell=40$, $K=6$, equal cluster sizes,
$\alpha_k^{(\ell)}=2$,
$(\phi_1,\phi_2,\phi_3)=(0.2,0.4,0.6)$ and innovation
standard deviations $(\gamma_1,\gamma_2,\gamma_3)
=(0.10,0.12,0.14)$. The slope templates, with clusters
in rows and conditions in columns, were
\[
    B=\begin{pmatrix}
    -2&-2&-2\\-2&-1&0\\0&-1&1\\
    0&1&-1\\2&1&0\\2&2&2
    \end{pmatrix}.
\]
A common slope multiplier was calibrated once so that
the minimum pairwise joint Mahalanobis distance between
the six reference mean profiles was four. It was then
held fixed across configurations. For
$r_i=(i-1/2)/n_\ell$, coordinates were
$\tau_i=\exp(10+0.4r_i^2)$, with standardized log
coordinates $z_i$ as in the mean model. Separate
configurations changed $p$ to 300 or 1,200,
$n_\ell$ to 20 or 80, or $K$ to three. The three-cluster
case used the first three slope templates with equal
cluster sizes. Thus, comparisons between configurations
describe their computational costs without holding
statistical difficulty constant.

Three further configurations retained the reference
dimensions but imposed the corresponding generating and
fitted covariance constraints. S1 used $\phi_\ell=0.4$
and $\gamma_\ell=0.12$; S2 used a common $\phi=0.4$
with the reference condition-specific standard deviations;
and S3 used the reference coefficients with
$\gamma_\ell=0.12$. These runs checked the implementation
under sharing constraints; they did not compare model
selection or the suitability of S1--S4.

For each dataset, three $K$-means partitions of the
observed concatenated profiles supplied common numerical
initializations. The retained fit maximized the
observed-data likelihood over these starts. All four
implementations shared the E-step, GLS mean updates,
$|\phi_\ell|\leq0.98$, innovation-variance floor
$10^{-8}$ and mean-system regularization $10^{-8}$.
The outer tolerance was $10^{-8}$ in log likelihood per
feature; the inner tolerance was $10^{-8}$ in the largest
parameter change, with maximum budgets of 1,000 outer
and 100 inner iterations. These benchmark settings
override the application defaults in the Method section.
They were frozen after checks on separate development
datasets using tighter tolerances and doubled iteration
budgets.

The blockwise numerical comparator used L-BFGS-B with
analytic gradients, curvature scaling and favorable
warm starts, replacing the explicit covariance updates
while retaining their block order. A second comparator
optimized all covariance parameters jointly. The
raw-curve comparator retained the closed updates and
shared iteration driver but evaluated likelihoods,
GLS systems and residual statistics through the direct
Python kernels. It was not a call to the upstream R
functions. Under S2, the shared coefficient and
condition-specific variances were alternated to the
inner convergence criterion.

\begin{table}[t]
    \centering
    \caption{Complete-fitting speedups under the common
    model. Each row has 30 independent paired datasets.
    Entries are median per-dataset comparator/C time
    ratios; brackets give 95\% bootstrap intervals for
    the primary N/C comparison. C denotes closed updates
    with cached statistics; N, blockwise numerical
    covariance optimization; J, joint numerical covariance
    optimization; and R, closed updates through direct
    raw-curve Python kernels. Configuration labels
    correspond to Fig.~\ref{fig:matched-speed-distributions}.}
    \label{tab:matched-runtime}
    \small
    \setlength{\tabcolsep}{2pt}
    \begin{tabular}{@{}lrrr@{}}
        \toprule
        Configuration & N/C [95\% CI] & J/C & R/C \\
        \midrule
        base & 2.19 [2.15--2.21] & 1.83 & 2.09 \\
        p300 & 2.37 [2.35--2.39] & 1.98 & 1.89 \\
        p1200 & 1.87 [1.83--1.97] & 1.60 & 2.40 \\
        n20 & 2.21 [2.10--2.26] & 1.86 & 1.80 \\
        n80 & 2.01 [1.84--2.07] & 1.68 & 2.66 \\
        K3 & 2.17 [2.03--2.33] & 1.86 & 1.93 \\
        S1 & 2.04 [2.00--2.07] & 1.81 & 2.06 \\
        S2 & 2.06 [2.03--2.10] & 1.86 & 2.04 \\
        S3 & 2.13 [2.12--2.18] & 1.85 & 2.11 \\
        \botrule
    \end{tabular}
\end{table}

\begin{figure*}[hbp]
    \centering
    \includegraphics[width=\linewidth]{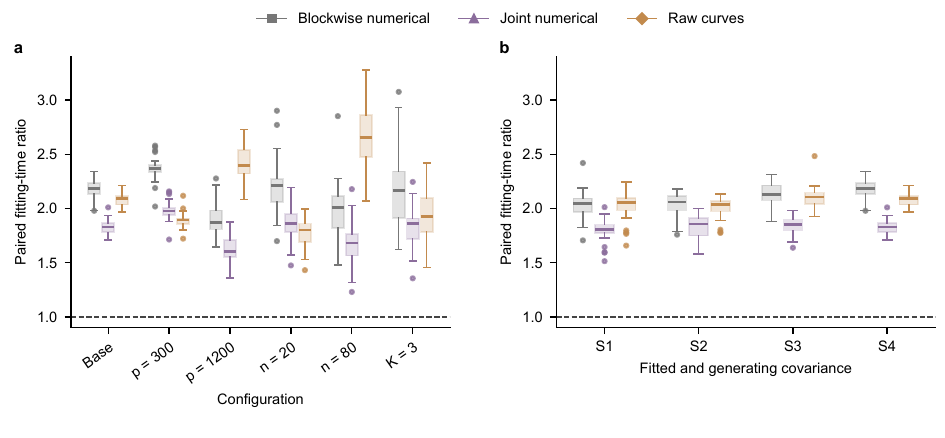}
    \caption{Distributions of paired computational gains.
    (a) Comparator/closed-and-cached fitting-time ratios
    across size configurations. Base has $L=3$, $p=600$,
    $n_\ell=40$, $K=6$ and S4 covariance; other settings
    change the indicated quantity. The three-cluster
    case uses the first three slope templates.
    (b) Ratios under correctly specified S1--S4
    constraints at reference dimensions. Every box
    contains 30 independent paired-dataset ratios.
    Boxes span the quartiles, center lines mark medians,
    whiskers reach the most extreme observations within
    1.5 interquartile ranges, and circles show all
    observations beyond the whiskers. The dashed line
    denotes equal fitting time. The Base and S4 groups
    use the same reference datasets. These sharing
    comparisons evaluate computational performance
    under each specification.}
    \label{fig:matched-speed-distributions}
\end{figure*}

Fixed-mean, fixed-weight covariance comparisons were
also made within this common Python backend, using ten
datasets for each S specification and states after
initialization and after up to five outer iterations.
Blockwise numerical/closed time ratios were
15.49--23.81, with maximum normalized conditional
objective difference $6.66\times10^{-16}$. These timings
excluded construction of the quadratic statistics;
complete-fitting timings included backend preparation.
This distinction also separates these measurements
from the R comparisons below.

No common-model fits were invalid or incomplete.
There were 557 non-success optimizer statuses in
296,092 blockwise numerical calls and 130 in 148,053
joint numerical calls. These were conditional optimizer
diagnostics, rather than failed complete fits. No
numerical updates were rejected for objective
deterioration, and all final likelihood-agreement checks
passed. Agreement between implementations establishes
their relative behavior under the common three-start
budget, rather than attainment of a global mixture
likelihood maximum.

\subsection{Upstream software and generating models}

We used the unmodified FunClu source files from
\texttt{cxzdsa2332/\allowbreak{}idopNetwork}, version~0.1.2,
at commit \texttt{a1a00a2789\allowbreak{}0b5c9d09e9\allowbreak{}9140c9284d\allowbreak{}37c11e2f2f}
(source repository: \url{https://github.com/cxzdsa2332/idopNetwork}).
The functions \texttt{fun\_clu} and
\texttt{bifun\_clu} fit power-law means and jointly
optimize mean and covariance parameters numerically.
The single-condition function uses BFGS. In the
two-condition function, nontrivial bounds cause R's
\texttt{optim} to select L-BFGS-B, although the source
passes the method name BFGS. The original stopping
rules and maximum of 100 outer iterations were retained.
The current implementation used the common-model
benchmark tolerances above and S4 covariance blocks.

For both mean families, let $t_i$ be equally spaced on
$[1,2]$ and $z_i$ its standardized value, corresponding
to $\tau_i=\exp(t_i)$ in the current mean model.
For zero-based condition index $h=\ell-1$, define
$a_{k\ell}=a_k+0.2h$ and
$b_{k\ell}=b_k+0.1h$, where
\[
\begin{gathered}
    (a_1,\ldots,a_6)=(3,3.6,4.2,4.8,5.4,6),\\
    (b_1,\ldots,b_6)=(0.4,0.6,0.8,0.5,0.7,0.9).
\end{gathered}
\]
Linear means were $a_{k\ell}+b_{k\ell}z_i$;
power-law means were $a_{k\ell}t_i^{b_{k\ell}+0.1h}$.
The first $K$ templates were used with equal class
sizes. Independent finite-start SAD(1) residuals had
$\phi_\ell=(0.3,0.5)$ and
$\gamma_\ell=(0.12,0.15)$, taking the first pair for
single-condition tasks. The workflows received the
same responses on the modeled scale. Columns and
coordinates were aligned with the upstream column-sum
sorting rule before fitting.

The baseline tasks used $(p,n_\ell,K)=(300,20,3)$
with one or two conditions. Larger tasks used
$(600,40,6)$ with two conditions. Linear and power-law
generation gave six configurations. Four baseline
configurations each used 30 independent datasets;
the two larger configurations each used ten. The
repetition allocation and per-start upstream time
limit of 630 seconds were fixed after development,
before formal evaluation. Three common $K$-means
label starts were supplied to both methods through
their observed-data initialization interfaces;
mean coefficients and covariance starts were then
computed for the respective mean families.

The original returned fields and package scores were
retained. For consistent evaluation and start selection,
we independently calculated the observed-data likelihood
from the returned mean parameters, covariance parameters
and mixture weights. Final optimizer parameters were
scored separately, without replacing the upstream
returned values. Dense Gaussian replay agreed with
the innovation-form scoring to a maximum absolute
log-likelihood difference of $1.82\times10^{-11}$.
There were no nonzero optimizer statuses, outer-budget
limits or start-log errors in the formal upstream runs;
all current fits were valid.

\begin{table}[t]
    \centering
    \caption{Workflow costs relative to upstream FunClu.
    Times are medians in seconds, including common
    $K$-means and model-specific initialization and all
    three fits. Speedups are medians of paired ratios,
    so need not equal ratios of the displayed median
    times. Brackets give 95\% paired bootstrap intervals.
    Process startup, source loading and input reading
    are excluded.}
    \label{tab:upstream-workflows}
    \small
    \setlength{\tabcolsep}{2pt}
    \begin{tabular}{@{}llrrrl@{}}
        \toprule
        Task & Generation & Datasets & \shortstack{Upstream\\(s)} & \shortstack{Current\\(s)} & \shortstack{Speedup\\{}[95\% CI]} \\
        \midrule
        Single & Linear & 30 & 2.026 & 0.0125 & 156 [143--174] \\
        Single & Power & 30 & 2.647 & 0.0125 & 211 [185--228] \\
        Joint & Linear & 30 & 8.144 & 0.0163 & 499 [468--509] \\
        Joint & Power & 30 & 8.037 & 0.0163 & 488 [467--515] \\
        Larger & Linear & 10 & 71.683 & 0.0233 & 2913 [2788--3227] \\
        Larger & Power & 10 & 68.748 & 0.0244 & 2811 [2340--3314] \\
        \botrule
    \end{tabular}
\end{table}

\subsection{Same-R conditional objectives and timing}

The covariance comparison used the first ten formal
datasets of each upstream configuration, selected by
index. Means came from the first observed-data power-law
initialization. Before forming the common posterior
weights, its covariance start was perturbed to
$\phi_{\mathrm{start}}=0.5\phi_{\mathrm{initial}}+0.15$
and $\gamma_{\mathrm{start}}=1.4\gamma_{\mathrm{initial}}$.
With these means and weights fixed, numerical solving
called the upstream \texttt{Q\_function} or
\texttt{biQ\_function} using BFGS with
\texttt{maxit=1000}, \texttt{reltol=1e-11} and
\texttt{parscale=0.1}. The objective was normalized by
$p$. Closed solving constructed the weighted residual
statistics and updated $\phi$ and $\gamma^2$ directly.
Both kernels used the same unconstrained conditional
coefficient domain. All resulting covariance parameters
had absolute values below 0.785, inside the upstream
two-condition bounds of $[-10,10]$. All 60 numerical
solves returned convergence code zero and met the
predefined $10^{-6}$ per-feature objective-agreement
criterion (Table~\ref{tab:same-r-runtime}).

\begin{figure*}[hbp]
    \centering
    \includegraphics[width=\linewidth]{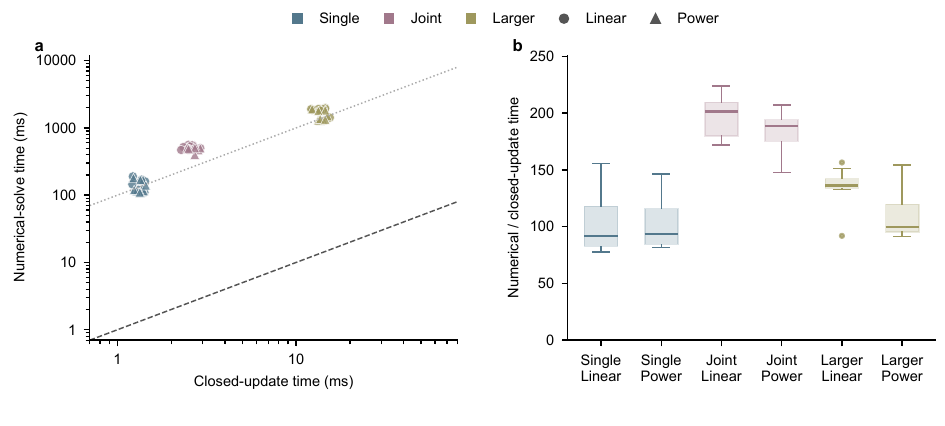}
    \caption{Covariance solving for the same conditional
    objective in R. Both implementations use fixed,
    identical power-law mean curves and posterior weights.
    Single and joint tasks have $(p,n_\ell,K)=(300,20,3)$
    with one and two conditions, respectively; larger
    tasks have $(600,40,6)$ with two conditions. Linear
    and Power identify the generating mean family.
    Each configuration uses ten independent datasets.
    (a) Paired numerical and closed solving times for
    all 60 targets, with task encoded by color and
    generating family by symbol. Both axes are
    logarithmic; dashed and dotted lines denote $y=x$
    and $y=100x$. The closed kernel includes weighted
    residual-statistic construction. (b) Per-dataset
    numerical/closed time ratios. Boxes span the quartiles,
    center lines mark medians, whiskers reach the most
    extreme observations within 1.5 interquartile ranges,
    and circles show all observations beyond the whiskers.
    These show empirical variability, not confidence
    intervals. Technical timing repetitions do not
    increase the number of independent datasets.}
    \label{fig:same-r-covariance}
\end{figure*}

\begin{table}[t]
    \centering
    \caption{Same-R conditional covariance comparisons.
    Each row uses ten independent datasets. Time ratios
    are numerical/closed, with 95\% paired bootstrap
    intervals. The last column is the maximum absolute
    conditional objective difference per feature.}
    \label{tab:same-r-runtime}
    \small
    \setlength{\tabcolsep}{2pt}
    \begin{tabular}{@{}llrl@{}}
        \toprule
        Task & Generation & \shortstack{Speedup\\{}[95\% CI]} & \shortstack{Maximum\\objective gap} \\
        \midrule
        Single & Linear & 91.5 [81.8--119.6] & $1.52\times10^{-10}$ \\
        Single & Power & 93.7 [82.5--122.2] & $1.58\times10^{-10}$ \\
        Joint & Linear & 201.4 [177.0--215.3] & $8.25\times10^{-11}$ \\
        Joint & Power & 188.8 [171.9--196.0] & $8.30\times10^{-10}$ \\
        Larger & Linear & 136.3 [133.4--143.8] & $1.31\times10^{-8}$ \\
        Larger & Power & 99.9 [94.6--125.8] & $7.66\times10^{-9}$ \\
        \botrule
    \end{tabular}
\end{table}

All timings were sequential, on the same Windows
machine with a single BLAS thread, with method order
randomized within datasets. Python~3.12.10 and
R~4.6.1 were used, with \texttt{mvtnorm}~1.4-2 for
upstream Gaussian densities. Core fitting time included
method-specific preparation and all three starts.
Initialization was recorded separately and added for
workflow timing. Conditional kernels were repeated
in groups lasting at least 0.2 seconds in Python
and 0.15 seconds in R; repetitions reduced timer
noise but did not increase the independent sample size.
All reported speedups are medians of per-dataset
ratios. Confidence intervals use 5,000 paired
dataset-level bootstrap resamples \citep{Efron1979}.
Development and
formal random streams were separate, and no formal
datasets were removed for timing or quality reasons.

\section{Additional real-data analysis details}
\label{app:idop-data}

The idopNetwork examples were obtained from the public
source revision used in the software benchmarks,
commit \texttt{a1a00a2}. The distributed
\texttt{gut\_microbe} object contained 65 rows and 20
sample columns; \texttt{mustard\_microbe} contained
3,263 rows and 176 sample columns. The latter differed
from the plant dimensions stated in the article and
package documentation. Matching the suffix-stripped
leaf and root column identifiers yielded 88 complete
sample pairs.

For the gut analysis, we used the 48-feature, 20-sample
input recovered from the package's saved
\texttt{test\_result} object, rather than rerunning its
random-start power-function screening. The saved
power-fitting and clustering objects contained identical
abundance matrices. Every retained abundance value and
sample identifier matched the distributed raw table.
The retained feature set equalled the set with at least
15 nonzero measurements out of 20. The ``Others'' row
was retained and identified as an aggregate feature.
For the plant analysis, the package cleaning function's
implemented threshold of at least 50 nonzero measurements
across 176 columns retained 3,137 OTUs.

For the human multi-omics and plant applications,
features were standardized within condition using their
sample mean and population standard deviation, followed
by a condition-wide shift making the minimum value one.
The response was the logarithm of these shifted values,
and their sample-level sum defined the niche index.
For the gut application, the response was
$\log_{10}(1+x)$ without feature standardization. Its
niche index was one plus the sum of retained raw abundances;
the base-10 logarithm of this index matched the coordinate
stored in the author's clustering object. The retained
gut matrix contained 90 zeros. In all applications,
the standardized logarithm of the niche index defined
the mean-function coordinate.

Each candidate used five initializations, seed 20261102,
a maximum of 1,000 outer iterations, convergence
tolerance $10^{-5}$ and five inner conditional updates.
S4 was fixed during cluster-number selection; for a
single condition, S1--S4 define the same covariance
family. The gut and plant scans covered $K=1,\ldots,20$
and $K=1,\ldots,40$, respectively.

For the gut analysis, a fitted candidate was admissible
only if each component had at least two features assigned
to it by maximum posterior probability. Empty components
counted as size zero. This requirement filtered the
candidate fits before BIC selection; it did not change
the EM updates or merge fitted modules. It was introduced
after inspection of the unconstrained results. The
$K=1,\ldots,7$ fits satisfied the requirement; the
$K=8,\ldots,20$ fits did not. Seven modules minimized
BIC among admissible fits. The unconstrained minimum
was eight modules (BIC $=2,191.92$), including two
singletons, compared with BIC $=2,198.39$ for seven modules.
This selection rule does not exclude the possibility
of another local solution at a rejected $K$ satisfying
the size requirement. In the selected fit, M6 contains
\textit{Escherichia coli} and ``Others''; it is not a
module of two individual species. Membership probabilities
summarize uncertainty within the selected fit.

For the plant visualization, each member profile was
regressed on the standardized log niche coordinate using
ordinary least squares. Member and mixture-center
functions were evaluated at 160 points within each
condition's observed coordinate range, then
back-transformed to the shifted standardized response.
Each module is displayed as a leaf--root panel pair.
For the gut visualization, all 960 observed
feature-by-sample values are shown as hollow circles,
with the component centers evaluated at 160 grid points.
Both observations and centers remain directly on the
modeled $\log_{10}(1+x)$ scale without back-transformation.
No member-function sampling, jitter or point subsampling
was applied to the gut display. Module panels within
each application share vertical limits and tick locations;
the BIC panel retains its own vertical scale.

As preprocessing sensitivity analyses on the same
48-feature gut input, we also examined feature-wise
Z-scores followed by a global shift to minimum one and
a natural logarithm, and feature-wise min--max scaling
to $[0,1]$ without a subsequent response transformation.
Both scans covered $K=1,\ldots,8$ without the minimum-size
filter, with niche indices recomputed from their respective
transformed inputs. The Z-score pipeline selected one
component (BIC $=31.20$), with $K=2$ higher by 11.61.
The min--max pipeline also had its lowest BIC at one
component, but $K=2$ was only 0.20 higher. These
transformations remove between-feature abundance-level
and amplitude differences, changing the profiles being
clustered. BIC values were compared across $K$ within
each response scale, not between response scales.

\section{Clustering recovery: design and additional results}
\label{app:clustering-quality}

\subsection{Generating model and controlled comparisons}

Each feature was represented by a 60-dimensional curve,
with $L=3$ conditions and $n_\ell=20$ positions per
condition. For condition $\ell=1,2,3$, we set
$u_{\ell i}=[1+(i-1)/19]^{1+0.12(\ell-1)}$
and standardized these values to obtain $z_{\ell i}$.
The supplied niche coordinate was $\tau_{\ell i}=\exp(u_{\ell i})$;
it was fixed independently of the simulated responses.
The four initial linear centers were
$a_{k\ell}+b_{k\ell}z_{\ell i}$, with
\[
A=\begin{pmatrix}
-0.5&0&0.35\\0.45&-0.45&0\\0&0.5&-0.45\\0.05&-0.05&0.1
\end{pmatrix},\qquad
B=\begin{pmatrix}
0.25&-0.35&0.1\\-0.3&0.2&0.35\\0.3&0.1&-0.25\\-0.25&0.05&-0.2
\end{pmatrix}.
\]
At each condition and position, we centered these four
means across modules, scaled the resulting mean array
to an average pairwise squared separation of $0.45^2$
over all conditions and positions, and added a common
intercept of 2. Training and test labels were balanced
across the four modules and randomly permuted.

Residuals followed the SAD(1) innovation model in the
Methods, independently across curves and conditions.
The coefficient $\phi$ was common to the three conditions
in generation; the fitted S4 model estimated separate
parameters for each condition. We chose the innovation
variance so that $\operatorname{tr}(\Sigma_\ell)/20=1$
for each $\phi$. Thus, changing correlation did not
change average marginal noise variance. The linear
design crossed $p\in\{300,600,1200\}$ with
$\phi\in\{0,0.4,0.8\}$.

The curved-mean control used $p=600$ at the same three
correlation levels. For each condition, we projected
$z_{\ell i}^2$ off the intercept and linear coordinate,
then normalized the remaining quadratic vector to unit
root mean square. We added this vector to each initial
center with coefficients $0.5c_{k\ell}$, where
\[
C=\begin{pmatrix}
1&-1&0.5\\-1&0.5&1\\0.5&1&-1\\-0.5&-0.5&-0.5
\end{pmatrix},
\]
before applying the same centering and separation scaling.
There were 30 independent datasets per setting, giving
360 datasets overall. Each had 2,000 independent test
curves at the same fixed coordinate positions. Responses
were generated directly on the modeled scale.

\subsection{Fitting and evaluation}

All methods used the same five $K$-means initial
partitions. $K$-means retained the fit with the smallest
within-cluster sum of squares; mixture methods retained
the fit with the largest final likelihood. The GMM
estimated unrestricted 60-dimensional mean vectors and
one shared full covariance matrix. It therefore modeled
correlation, including cross-condition covariance.
FunClu estimated condition-specific linear mean
coefficients and S4 SAD(1) covariance shared across modules.
The independent-residual variant used the same mean and
EM implementation, fixed $\phi_\ell=0$, and re-estimated
each residual variance from weighted residual squares.
The respective parameter counts were 2,073 for the GMM,
33 for FunClu and 30 for independent-residual FunClu.
The comparison measures the performance of these fitted
models jointly; it does not isolate parameter count
from all other differences in estimation.

Mixture fits used tolerance $10^{-6}$, at most 1,000
outer iterations and regularization $10^{-6}$; FunClu
used at most five inner iterations and $|\phi_\ell|\leq0.98$.
We matched estimated and true centers by minimizing
the total squared distance and computed RMSE across
modules, conditions and positions. Test ARI compared
predicted and true labels for new feature curves.
Bootstrap intervals used 5,000 resamples of independent
datasets; paired contrasts resampled the same datasets
for both methods. The generating-parameter classification
reference used the true means, covariance and weights.
It indicates intrinsic classification difficulty and
is not a finite-sample upper bound on ARI.

All 360 datasets and 1,440 selected fits were retained.
Both FunClu variants met their convergence criterion
in all datasets; the GMM did so in 359 of 360.
For its remaining fit (linear means, $p=1200$,
$\phi=0.4$, replicate 10), increasing the limit to
4,000 yielded convergence at iteration 1,037 and changed
test ARI from 0.09493 to 0.09624. The primary summaries
retain the original iteration limit. Formal simulations
used seed 20261103, separate from pilot random streams,
with Python 3.12.10, NumPy 2.5.1, SciPy 1.18.0
and scikit-learn 1.9.0.

\begin{figure*}[!htbp]
    \centering
    \includegraphics[width=0.92\linewidth]{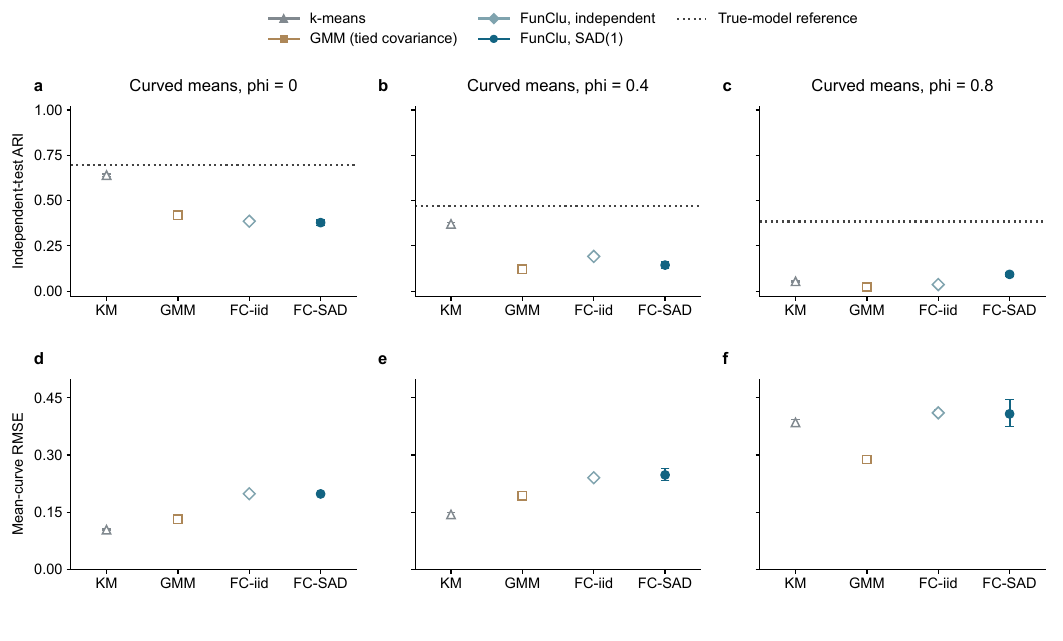}
    \caption{Recovery under curved generating means at
    $p=600$. Panels show independent-test ARI (top) and
    mean-curve RMSE (bottom) at three residual correlations.
    Points are means over 30 independent datasets;
    intervals are 95\% dataset-level bootstrap intervals
    (5,000 resamples). KM, GMM, FC-iid and FC-SAD denote
    $K$-means, the shared-full-covariance GMM,
    independent-residual FunClu and SAD(1) FunClu.
    The same separation and average marginal noise
    variance are used as in the linear design.}
    \label{fig:curved-mean-control}
\end{figure*}

\begin{figure*}[!htbp]
    \centering
    \includegraphics[width=0.92\linewidth]{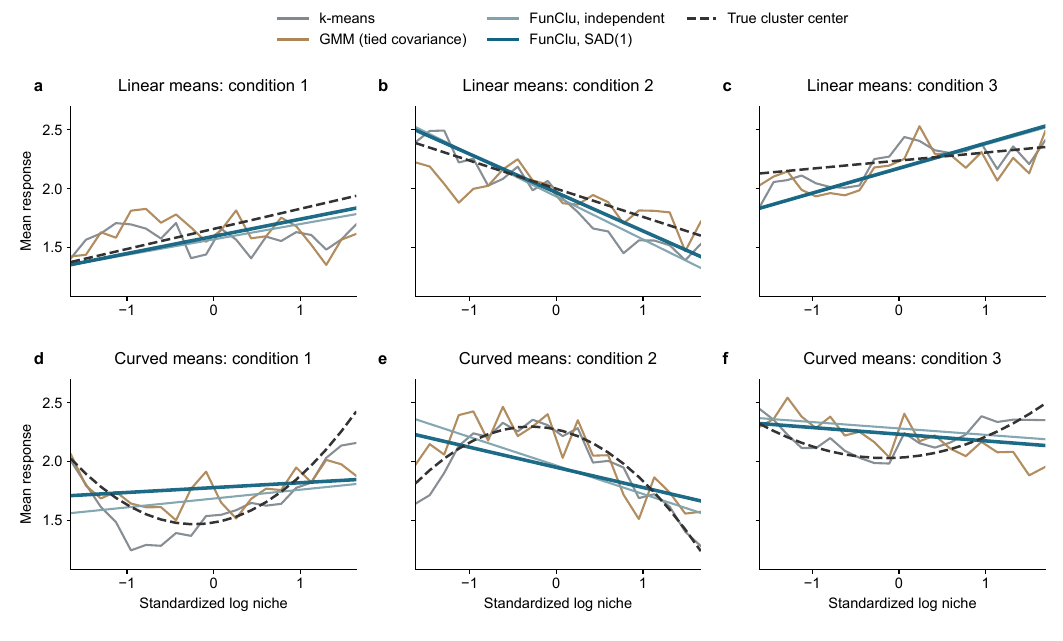}
    \caption{Estimated mean curves for a fixed example:
    the first formal replicate, true module 1,
    $p=600$ and $\phi=0.4$. Rows show linear and
    curved generating means; columns show the three
    conditions. Dashed curves are the true centers;
    colored curves are label-matched estimates.
    The example was fixed by replicate and module index.}
    \label{fig:quality-example-centers}
\end{figure*}

\end{appendices}
\clearpage

\bibliographystyle{oup-abbrvnat}
\bibliography{references}

\end{document}